\documentclass{iopart}  
\usepackage{epsfig}

\newcommand{\be}[1]{\begin{equation}\label{#1}}
\newcommand{\ee}{\end{equation}}   
\newcommand{\eq}[1]{(\ref{#1})}

\begin{document}

\letter{Dynamical stabilization of classical multi electron targets
against autoionization}

\author{Tiham\'{e}r Geyer\dag %
	\ and Jan M Rost\ddag}
\address{\dag Department of Chemical Physics, Weizmann Institute of Science, 
Rehovot 76100, Israel}
\address{\ddag Max--Planck--Institute for the Physics of Complex
Systems, N\"othnitzer Str. 38, D--01187 Dresden, Germany}

\begin{abstract}
	We demonstrate that a recently published quasiclassical M\o{}ller
	type approach [Geyer and Rost 2002, \emph{J. Phys. B} \textbf{35}
	1479] can be used to overcome the problem of autoionization, which
	arises in classical trajectory calculations for many electron
	targets. In this method the target is stabilized dynamically by a
	backward--forward propagation scheme. We illustrate this
	refocusing and present total cross sections for single and double
	ionization of helium by electron impact.
\end{abstract}


\submitto{\jpb}

\vspace{5mm}



Classical trajectory descriptions of atomic collisions and ionization
processes have a long history, dating back to the pioneering work of
Abrines and Percival \cite{ABR66}. The method has never become a
mainstream tool, but has been used over the years for a variety of
collision systems. This so called CTMC (Classical Trajectory
Monte--Carlo) method was originally formulated with macroscopic point
particles, scaled down to the dimensions of a real hydrogen atom, but
it can be derived as a discretized treatment of the system's Liouville
equation, too (see, e.g. \cite{KEL93}). In the hydrogen target, for
which CTMC was formulated initially, the single electron orbits around
the nucleus on a Kepler ellipse. If this concept is extended, the
resulting many electron atom is highly unstable: the electrons
exchange energy and finally one of them ends up in a very tightly
bound orbit and all the others are kicked out of the atom. This
process is called autoionization, as it does not need any external
perturbation. Therefore it is practically impossible to use such a
classical many electron atom as a target in a CTMC calculation --- it
dissolves on its own before the approaching projectile has any chance
to interact with it.

Various attempts have been made to stabilize classical many electron
atoms, so that they could be used in ionization and excitation
calculations. These attempts range from neglecting the target
electrons' interaction completely in the independent electron
model, through highly symmetric initial configurations, which
autoionize slightly slower \cite{SCH92}, up to additional momentum
dependent potentials to incorporate the uncertainty relation
\cite{KIR80}. These ans\"{a}tze are then usable in CTMC calculations,
but they either describe a modified scattering system or lead to
inconsistencies.

Motivated by these and other shortcomings of the standard CTMC method
we recently proposed a quasiclassical description for particle impact
ionization, which is derived as an approximation to the quantum
mechanical description \cite{GEY02}: the M\o{}ller formulation of the
quantum scattering operator is translated into the Wigner phase space
formalism \cite{WIG32} and finally approximated by setting $\hbar \to
0$. The approximation procedure itself is well established and the
resulting method is technically very similar to CTMC, as in both
treatments the cross sections are evaluated by propagating classical
trajectories. But nevertheless there are two fundamental differences:
(i) the phase space description of the target's initial state is
derived free of ambiguity from the quantum mechanical wave function
without the need to artificially introduce quantization recipes for
many electron atoms and (ii) the M\o{}ller form of the scattering
operator translates into a classical backward--forward--backward
propagation scheme, which stabilizes arbitrary initial distributions.
With this ansatz we could calculate fully differential cross sections
for electron impact ionization of hydrogen, which essentially reproduce
the experimental results over a wide range of
energies and geometries \cite{GEY02}.
We will now demonstrate that this approximation can also deal with an
autoionizing classical helium target.

As mentioned above our quasiclassical approximation is derived as the
$\hbar=0$ limit of the Wigner formulation of the quantum scattering
operator $\hat{S}$ in the M\o{}ller form (for details, please
see \cite{GEY02} and references therein):
\[
	\hat{S} = \Omega^\dag_- \Omega_+ 
	\quad \mbox{with} \quad
	\Omega_\pm = \lim_{t \to \mp \infty} U^\dag(t) U_0(t)
\]
The propagators $U(t) = \exp [-\rmi Ht]$ and $U_0(t)$ finally
translate into solving Hamilton's equations of motion for each
of the (multi dimensional) discretization points of the initial
distribution
\[
	\rho(t=0) = \rho^i(x,p) = 
		\mathcal{N} \sum_n w_n \delta(x-x_n) \delta(p-p_n).
\]
The weights $w_n$ are the values of the $\hbar=0$ limit of the Wigner
transform $w^i$ of the initial state wave function at the
discretization point: $w_n = w^i(x_n, p_n)$. It can be shown that the 
Wigner transform is only one special case to select the initial 
conditions; by modifying the underlying 
correspondence rule nearly arbitrary translations between
wave functions and phase space distributions can be constructed 
\cite{MEH64, COH66}. We will later use this freedom to calculate 
cross sections with a more simple initial state distribution.

According to the M\o{}ller scheme each trajectory first is propagated
backward in time under the asymptotic initial Hamiltonian $H_0^i$,
i.e., with the interaction between target and projectile switched off.
If not denoted otherwise we will in the following use atomic units.
Its form is then:
\be{eq:HInitial}
	H_0^i = \frac{p_p^2}{2} + 
		\frac{p_1^2}{2} - \frac{Z}{r_1} + 
		\frac{p_2^2}{2} - \frac{Z}{r_2} + 
		\frac{1}{|r_1-r_2|}
\ee
The subscript $p$ denotes the projectile whereas the target electrons
are labeled with 1 and 2. The nucleus is set to have an infinite
mass.

When projectile and target are separated far enough, denoted
symbolically by $t=-\infty$, the interaction is switched on and the
trajectory evolves forward again under the full Hamiltonian $H$
through the collision at $t=0$ and on, until the fragments are well
separated again. Then the fragments are brought back from $t=\infty$
independently, i.e., with the asymptotic final $H^f_0$, to the initial
time $t=0$. If the initial state is unstable under the classical
propagation, as is the case with a helium target, then it autoionizes
already during the first backward propagation. When the
projectile--target interaction is added at the turning point
$t=-\infty$, it is negligible first; the forward propagation
effectively undoes the autoionization and the projectile encounters
the refocused target in nearly its initial state. The autoionization
still takes place, but now it is shifted away to $t<0$, where it has
no influence on the actual collision dynamics, which takes place
around $t=0$. There is consequently no need any more to neglect some
part of the interactions or to introduce additional stabilizing
potentials.

In a quantum treatment the first backward propagation only contributes
a phase shift, as the initial state is an eigenstate of $H^i_0$.
The cross sections remain unchanged, if it is neglected. In
the classical approximation, though, the target is not stationary
during the backward and the forward propagations; but if both are
performed, most of the error due to the approximation cancels and the
target is effectively stationary with respect to the central time 
$t=0$: it is this point in time, where the initial conditions are set 
up, where the collision takes place and where finally the cross 
sections are extracted.


\epsfxsize=11cm
\epsfysize=5cm
\begin{figure}[tp]
  \centerline{\epsfbox{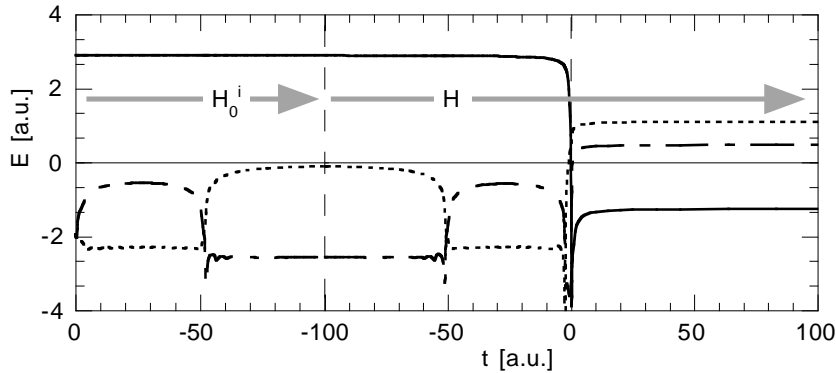}}
  \caption{One electron energies of the three electrons in the course
  of an example trajectory. The two target electrons (broken curves)
  each start at $-1.94$ a.u. The propagation starts backwards with
  $H_0^i$ \eq{eq:HInitial} until $t=-100$ a.u. and is then reversed.
  The forward propagation is performed with the full Hamiltonian $H$.
  The energy of the projectile (solid line) is shifted from $E_p = 2$
  keV to $E_p = IP_1+IP_2$. For further explanations please see text.}
  \label{fig:ETraj}
\end{figure}

The stabilizing effect of this M\o{}ller type
backward--forward--backward scheme is demonstrated in figure
\ref{fig:ETraj}: there the one-electron energies $E_n =
\frac{p_n^2}{2} - \frac{2}{r_n}$, $n=p, 1, 2$, of the projectile
(solid line) and the two target electrons (broken lines) during one
trajectory, i.e., the evolution of one single discretization point,
are plotted against the propagation time. The interaction energies
between the electrons are not included. The propagation starts at
$t=0$ and first proceeds backwards, here up to $t=-100$ a.u., under
$H_0^i$ \eq{eq:HInitial}, i.e., with target and projectile independent
of each other. Then the propagation reverses and the full $H$ is used
to propagate back to $t=0$ and through the scattering event. The plot
ends at ``$+\infty$'', which is $t=100$ a.u. here. In the final
backward propagation the electrons are independent of each other,
i.e., their energies do not change any more. Hence, we do not show 
this part of the trajectory in figure \ref{fig:ETraj}.

The target electrons were started
in a symmetric configuration for the trajectory of  figure \ref{fig:ETraj}.
Within less than one period one of them
is kicked into a large orbit, from which it returns after about 50 a.u.
and kicks out the other electron. After the propagation is reversed it
is clearly seen that for most of the now following forward propagation
the energies of the target electrons ``rewind'' the backward
propagation, i.e., the broken lines are symmetric with respect to
$t=-100$ a.u. In other cases (not shown here) one of the
electrons is even kicked out of the atom to a positive energy and only
comes back due to the reversed propagation. The projectile's energy of
$E_p$ = 2 keV = 73.5 a.u. is shifted in the plot by $E_p-(IP_1+IP_2)$,
i.e., the projectile is plotted to start at the negative energy of the
target. $E_p$ is constant during the backward and most of the forward
propagation. It only starts to change shortly before $t=0$, when the
interaction with the target electrons increases. Then the symmetry of the
target electron energy trajectories against $t=-100$ a.u. is broken
and all three electron--electron interactions together determine the
dynamics of the actual ionization event. In this plot both electrons
are lifted to positive energies, i.e., double ionization occurs.


Another difficulty in the classical description arises from the 
fact that the final state is normally of well defined energy. Since 
the Hamiltonian is conservative only that part of the initial 
Wigner distribution contributes which lies on this energy shell.
Consequently, the quantum distribution character of the initial state 
is lost. To overcome this problem we proposed in \cite{GEY02} to evaluate the cross
sections in terms of the energy transfer, which, for hydrogen targets,
is equivalent to looking at only the projectile's energy.

With the helium target a trajectory contributes to double ionization,
when the projectile's energy loss $\Delta E_p$ is bigger than the sum
of the ionization potentials $IP_1$ and $IP_2$ of the target
\emph{and} when both the target electrons have gained at least half of
the total binding energy:
\be{eq:DoubleCond}
	-\Delta E_p > IP_1 + IP_2 \quad \mbox{and} \quad
	\Delta E_1, \Delta E_2 > \frac{IP_1+IP_2}{2}
\ee
A contribution to single ionization is consequently defined by
\be{eq:SingleCond}
	-\Delta E_p > IP_1, \quad
	\Delta E_1 > \frac{IP_1+IP_2}{2} \quad \mbox{and} \quad
	\Delta E_2 < \frac{IP_1+IP_2}{2}.
\ee
Of course, this test has to be performed with the target electrons'
energy transfers $\Delta E_1$ and $\Delta E_2$ swapped, too.

To verify the conditions \eq{eq:DoubleCond} and
\eq{eq:SingleCond} for double and single ionization we need the
initial and the corresponding final state wave functions. The initial
state is the same in all cases. We are therefore, as in the
experiment, able to extract \emph{all} physically feasible cross
sections from the same set of final values of the propagated
trajectories.


\epsfxsize=8cm
\epsfysize=6cm
\begin{figure}[tp]
  \centerline{\epsfbox{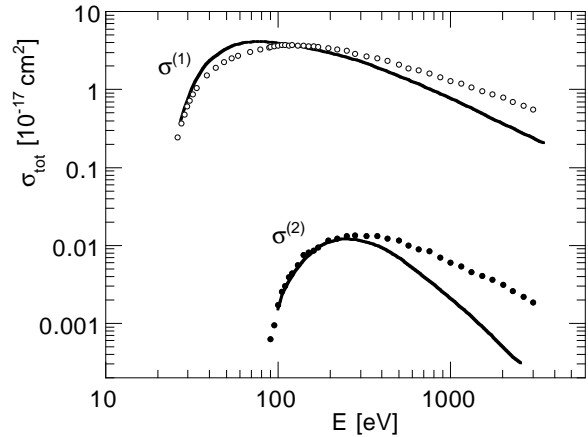}}
  \caption{Total cross sections $\sigma^{(1)}$ for single and
  $\sigma^{(2)}$ for double ionization: Comparison of our results
  (solid lines) with the experimental data of Shah \etal.
  \cite{SHA88} (open and filled circles).}
  \label{fig:Stot}
\end{figure}

As a first test of the performance and consistency of our approach
with the helium target we calculated the absolute total cross sections
$\sigma^{(1)}$ for single and $\sigma^{(2)}$ for double ionization.
They are compared to absolute measurements by Shah \etal \cite{SHA88}
in figure \ref{fig:Stot}.

For this first calculation we have used a  simple initial distribution 
obtained from a product wave function for the ground state of the helium
target
\be{eq:HeWF}
	\psi(r_1, r_2) = \frac{\tilde{Z}^3}{\pi} 
		\exp(-\tilde{Z}r_1) \exp(-\tilde{Z}r_2)
\ee
with effective nuclear charge $\tilde{Z}=\frac{27}{16}$ \cite{HYL29}.
Each of the single electron wave functions is then translated into a
phase space distribution by multiplying its densities in coordinate
and momentum space \cite{COH66}. Both the wave function \eq{eq:HeWF}
and the resulting phase space distribution have a total energy of
$E=-2.85$ a.u., slightly less than the experimental value of $-2.904$
a.u.

The single ionization cross section $\sigma^{(1)}$ reproduces the
measured data on the level of accuracy that is typical for a single
electron CTMC calculation, see, e.g., \cite{SCH92}: the maximum occurs
at a lower energy and is slightly higher than the experiment, while
the high energy behavior follows the classical $1/E$ decay
\cite{THO12}. The explicit treatment of both target electrons and of
all interactions can, of course, not reintroduce quantum effects like
tunneling. On the other hand the accessible phase space volume is much
bigger than with only one active electron and one might fear that the
dynamics ``strays away'' from the reaction path of single ionization,
completely distorting the cross section --- which obviously does not
happen. In fact our result is in good agreement with an nCTMC
calculation by Schultz \etal \cite{SCH92}.

The more interesting result is, of course, the double ionization cross
section $\sigma^{(2)}$: figure \ref{fig:Stot} shows the first
classical trajectory result ever, in which the dynamics according to
the correct full Hamiltonian without any modifications or additions
was solved -- simultaneously both for $\sigma^{(1)}$ and
$\sigma^{(2)}$.
The double ionization cross section has two regions of different
correspondence with the experiment: for high energies it decays as
$1/E^2$, much faster than the experimental data. This suggests that in
our calculation in the high energy regime both electrons are ionized
independently, each contributing a classical factor of $1/E$, and not
in a sequential event, which should decay approximately as
$\sigma^{(1)}$ \cite{REJ02}.
For impact energies below 250 eV, on the other hand, the experiment is
reproduced remarkably well, both in shape and in magnitude. In that
region no microscopic quantum mechanical explanation has been proposed
yet. It is known, though, that right above the threshold the final
state is defined only by the long range and long time dynamics of the
outgoing electrons, which can be well described classically
\cite{WAN53, PAT98}. The good agreement between our classical result
and the measured data shows that up to a total energy of about twice
the total binding energy of the helium target the main reaction paths
are the classical ones.

In this letter we have demonstrated for the first time that electron
impact ionization of a two electron atom, i.e., helium, can be
calculated within a classical trajectory Monte--Carlo framework with
the full, unmodified helium Hamiltonian. This has been achieved with a
quasiclassical M\o{}ller formalism: the propagation scheme refocuses
and stabilizes the autoionizing target. The total cross sections,
extracted from the energy transfer, compare well with the experiment
to within the limitations of the classical approximation.

The next, more demanding level of tests will be to compare the
differential cross sections to experimental results and finally to
understand the dynamics of double ionization in the low energy regime.

This work was funded by the Israel Science Foundation.


\end{document}